\documentclass[aps,prl,showpacs,superscriptaddress]{revtex4}
\usepackage{amsfonts,amsmath,graphicx}

\def \o{\omega}
\def \e{\varepsilon}
\def \be{\begin{equation}}
\def \ee{\end{equation}}

\begin{document}

\title{Beam propagation in finite size photonic crystals and metamaterials}
\author{B. Guizal}
\affiliation{ D\'{e}partement d'optique, Institut FEMTO-ST, UMR\ 6174\\
Universit\'{e} de Franche-Comt\'{e}\\
16, Route de Gray 25030 Besan\c{c}on Cedex \ \ France\\ }

\author{D. Felbacq}
\affiliation{Groupe d'Etude des Semiconducteurs \\
Unit\'{e} Mixte de Recherche du Centre National de la Recherche
Scientifique
 5650\\
Universit\'{e} Montpellier II\\
34095 Montpellier Cedex 5, France\\}

\author{R. Sma\^{a}li }
\affiliation{Laboratoire des technologies de la
micro\'{e}lectronique - 17 rue des Martyrs, 38054 Grenoble, Cedex 9
France.}
\date{\today }

\begin{abstract}
The recent interest in the imaging possibilities of photonic
crystals (superlensing, superprism, optical mirages etc...) call 
for a detailed analysis of beam propagation inside a
finite periodic structure. In this paper, an answer to the question "where does the beam emerge?" is given. Contrarily to common knowledge, it is not
always true that the shift of a beam is given by
the normal to the dispersion curve. This phenomenon is explained in
terms of evanescent waves and a renormalized diagram that gives the correct direction is given.
\end{abstract}
\maketitle
\section{Introduction and setting of the problem}

Some beautiful experiments and numerical works have shown that it
was possible to obtain quite unusual behaviors of light propagation
inside meta-materials and photonic crystals (PhCs)
\cite{ozbay,tret,fang,moi2,centeno,benisty,matsu,pendry}.
 In particular, the near field properties of meta-materials are intensively studied
 in view of the possibility of designing superlenses \cite{lens1,lens2} or cloaking devices \cite{cloak}.
 
  In these structures, the evanescent
waves play an important role: the point of this
work is to to quantify the importance of the evanescent waves and to give a theoretical insight into the propagation of a beam \cite{notomi} inside a finite-size
PhC.
In principle, the group velocity  \cite{fot,Luo} allows to determine where the beam emerges from the PhC (see figure 1) by computing the
shift $\Delta$.
We show that, in finite-size structures, the shift is not always correctly predicted by the normal to the isofrequency curves. 

This fact is due to two reasons. First, in finite-size structures, there are evanescent waves near the boundaries
which can contribute to the propagation of the beam \cite{moiprl}. Therefore the field inside the structure comprises not only Bloch modes but evanescent waves as well \cite{physicae}. The latter can have a strong influence on 
the behavior of the beam.
Second, due to multiple scattering, the emerging field is a sum of beams that can strongly interfere.
If the beams are well separated spatially, on can clearly distinguish where the first beam emerges from the structure. If the beams overlap strongly,
the resulting field can be strongly deformed, and it becomes difficult to define an "emerging point".

Throughout this work, we use time-harmonic fields, with a
time-dependence of $\exp (-i\o t)$. 
The fields are assumed to be $z$-independent (this is the direction of invariance of the photonic crystal). The vectorial diffraction problem can then be reduced to the study
of the two usual cases of polarization: $s$-polarization (electric
field parallel to $z$) or $p$-polarization
(magnetic field parallel to $z$). The wavenumber in the medium surrounding the photonic crystal is denoted
by $k_{0}$. The incident field is a limited beam whose $z$ component is given by
\begin{equation}
u^i \left( x,y\right) =\int_{-k_0}^{k_0} A\left( k_x \right) e^{i\left(  k_x x +k_{y0} y\right)} dk_x
\end{equation}
where: $k_{y0}=\sqrt{k_0^2-k_x ^2}$ and $A(k_x)$ is the spectral amplitude (for instance, it can be chosen gaussian:
$A\left( k_x \right) =e^{-\frac{w^{2}}{4}\left( k_x -k_{m}\right) ^{2}}$),
where $k_{m}=k_0 \sin \theta$, and $\theta$
 is the mean angle of incidence of the beam and $w$ its
waist.
The point where the incident beam enters the photonic crystal is defined as the barycenter of the beam. Its abscissa is given by: $X_{i}=\frac{\int x \left| u^{i}\left( x,0\right) \right| ^{2}dx}{\int \left| u^{i}\left( x,0\right) \right| ^{2}dx}$

\section{Shift of the first transmited beam}

The crystal is described as a stack of gratings (figure 1) and we assume that
in the spectral domain defined by the above beam, the ratio between
the wavelength and the period $d$ of the gratings is such that
there is only one reflected and one transmitted order
 (the condition $k_0 < \pi/d$ is sufficient).
Given this hypothesis, the reflected and transmitted
fields can be expressed as:
\begin{eqnarray}
u_{r}\left( x,y\right) =\int A\left( k_x \right) r_{N}\left( k_x
\right) e^{i\left( k_x x-k_{y0} y\right) }dk_x \\
u_{t}\left( x,y\right) =\int A\left( k_x \right) t_{N}\left( k_x
\right) e^{i\left( k_x x+k_{y0} y\right) }dk_x
\end{eqnarray}
where $r_{N}$ and $t_{N}$ are the reflection and transmission coefficients.
For a given $k_x$, there exists a unique real $2\times 2$ matrix
$\mathbf{T}_{N}$ \cite{pochiyeh,moi1} with determinant $1$, satisfying the following
relation:
\begin{equation}
\mathbf{T}_N\left(
\begin{array}{c}
1+r_{N} \\
ik_{y0}\left( 1-r_{N}\right)
\end{array}
\right) =t_{N}\left(
\begin{array}{c}
1 \\
ik_{y0}
\end{array}
\right)  \label{transfert}
\end{equation}
It is the dressed transfer matrix of the
total structure \cite{moiprl}. 
Let us denote by $\gamma$ and $\gamma^{-1}$ the eigenvalues of $\mathbf{T}_N$ and by
$\mathbf{v}=\left( \phi_{11},\phi _{21}\right) ,\mathbf{w}=\left( \phi _{12},\phi _{22}\right)$
the associated eigenvectors ($\mathbf{T}\mathbf{v}=\gamma \mathbf{v}\, ,\,
\mathbf{T}_N\mathbf{w}=\gamma^{-1} \mathbf{w}$).
The reflection and transmission coefficients can be written in the following form:
\begin{equation}
r_{N}=\frac{\left( \gamma^2-1\right) f}
{ \gamma^2-g^{-1}f}\, ,\,\,\, t_{N}=
\frac{ \gamma \left( 1-g^{-1}f\right) }
{ \gamma^2-g^{-1}f}
\end{equation}
where, denoting $q(x,y)=(ik_{y0}y-x)/(ik_{y0}y+x)$,
the functions $f$ and $g$ are defined by $g\left( k,\theta \right)
=q\left( \mathbf{v}\right) ,f\left( k,\theta \right) =q\left(
\mathbf{w}\right)$ and $\mathbf{v}$ is chosen such that $|g|\leq1$
in the conduction bands. 

The following expansions hold:
\begin{eqnarray}
r_{N}\left( k,\theta \right)  &=&g+\left( g-f\right) \sum_{p=1}^{+\infty
}\gamma ^{2p}\left| g\right| ^{2p} \\
t_{N}\left( k,\theta \right)  &=&(1-|g|^2)\gamma \sum_{p=0}^{+\infty }\gamma ^{2p}\left| g\right| ^{2p}
\end{eqnarray}
These series expansion show that, due to the multiple scattering inside the photonic crystal, the transmitted (and reflected) fields consist of a sum of beams $u_t(x,y)= \sum_{p=0}^{+\infty }u_t^p$, where:
\be
u_t^p(x,y)=\int A(k_x) (1-|g|^2)\left| g\right| ^{2p}\gamma ^{2p+1}
e^{i\left( k_x x+k_{y0} y\right) }dk_x
\ee
The position where a beam emerges from the
photonic crystal is defined as its barycenter. 
We are interested in the direction that is followed by the beam inside the structure. This direction is given by the shift of the first transmitted beam (see figure 1), the shift being defined as the difference between the barycenter of the incident field and that of the first transmitted field:
$\Delta =X_t-X_i$, where: $X_t=\frac{\int x \left| u^0_t\left( x,0\right) \right| ^{2}dx}{\int \left| u^0_t\left( x,0\right) \right| ^{2}dx}$.

We assume that for the considered frequency all the plane waves inside the beam belong to the conduction band of the photonic crystal. This implies that $\left|\gamma \right|=1$, it can therefore be written 
under the form: $\gamma=e^{ik_y\,Nh}$. The  vector $(k_x,k_y)$ is a "renormalized" Bloch vector in the following sense:
denoting by $u(x,0)$ the value of the field on the lower interface, we have: 
$u(x,Nh)=e^{ik_x\, x}e^{ik_y\, Nh}u(x,0)$.  
		
 Let us concentrate on the first transmitted beam, i.e.
the beam that reads:
\begin{equation}
u_{0}^{t}\left( x,Nh\right) =\int A ( k_x ) (1-\left| g\right| ^{2} ) e^{ik_y Nh}e^{ik_x x}dk_x
\end{equation}


Using Parseval-Plancherel identity, we get the angular shift due to the beam propagation  (cf.
fig. 1):
\begin{equation}\label{shift}
\frac{\Delta}{Nh}=-\frac{\int \frac{dk_y}{dk_x} A^2\left( k_x \right) \left( 1-\left|g\right| ^{2}\right)^2   dk_x }
{\int A^2\left( k_x \right) \left( 1-\left|g\right| ^{2} \right)^2 dk_x }
\end{equation}

A series expansion of $\Delta$ can be obtained provided the phase function is analytic with respect to
$k_x $ in a neighborhood of $k_{m}$. Then we obtain:

\begin{equation}\label{tanpsi}
\Delta=-Nh\sum_{m=0}^{+\infty}
\frac{C_m}{m!}
\left.\frac{d^{m+1} k_y}{d k_x^{m+1}}\right|_{k_{x0}}
\end{equation}
where
$C_m=\frac{\int A^2 ( k_x ) (1-\left| g\right| ^{2} )^2
\left( k_x -k_{x0}\right) ^{m}
 dk_x }{\int A^2 ( k_x ) (1-\left| g\right| ^{2} )^2 dk_x }$.

When $ A\left( k_x\right) $ is concentrated around $k_{m}$, and if
$\frac{d k_y}{d k_x}$ does not vary
too quickly in the vicinity of $k_{m}$, we obtain the
well-know crude approximation:
\begin{equation}
\Delta  \sim -Nh\frac{d k_y}{d k_x}\left(k_{m}\right)  \label{gracu}
\end{equation}

In order to give a geometric interpretation of this result, let us remark
that $\left( \frac{d k_y}{d k_x} \left( k_{m} \right) ,-1\right) $ is a
vector that is normal to the dispersion curve at wavelength $\lambda $. 

We retrieve the well-known fact that for a spatially large beam, the direction of
propagation is given by the normal to the isofrequency Bloch diagram \cite{iso} but here
\textbf{it is the "renormalized" Bloch diagram that is involved}. 

We will see in the numerical
application that it can be quite different from the usual Bloch diagram.

In the particular case of a one dimensional stratified medium, i.e. when the
relative permittivity is constant in the horizontal direction, 
the direction of propagation of a beam is given by the normal to the usual dispersion curve.
Indeed, denoting $\mathbf{T}$ the transfer matrix for $1$ period, the transfer matrix for $N$ periods is $\mathbf{T}^N$. This matrix coincides with the matrix $\mathbf{T}_N$ : this is due to the absence of evanescent waves. A direct consequence is that
the Bloch vectors obtained from $\mathbf{T}^N$ and $\mathbf{T}_N$ are the same, hence the renormalized
Bloch diagram coincides with the non-renormalized one. 

\section{Discussion}

In the following, we present numerical computations
illustrating the behavior of a beam inside finite PhCs. All the numerical results were obtained
by means of a rigourous diffraction code for gratings based on the Fourier
Modal Method (see \cite{FMM}).
We will denote:
\begin{itemize}
\item $\Delta_n$ the shift of the first transmitted beam computed by a direct numerical computation of the
field.
\item $\Delta_{B}$ the shift computed through the isofrequency
 Bloch diagram.
\item $\Delta_{R}$ the shift computed through the renormalized isofrequency diagram
\item $\Delta_{f}$ the shift of the entire transmitted field comprising all the beams, computed numerically.
\end{itemize}

In order to quantify the role of the evanescent waves, let us remark that, inside the 
photonic crystal, the field can be expanded over three types of modes \cite{physicae}. These modes are
the eigenvectors of the transfer matrix ${\cal T}$ of the photonic crystal, i.e. the matrix that relates the fields  below the crystal to the fields above the crystal:
\begin{enumerate}
\item the propagative modes, i.e. the Bloch modes (corresponding to the eigenvalues of ${\cal T}$  of modulus $1$),
\item the evanescent modes (corresponding to the eigenvalues of ${\cal T}$  of modulus less than $1$),
\item the anti-evanescent modes (corresponding to the eigenvalues of ${\cal T}$  of modulus greater than $1$).
\end{enumerate}
By projecting the field on these modes, it is possible to compute the
ratio of the field that is carried by the evanescent, anti-evanescent and propagating waves \cite{physicae}.
We consider a photonic crystal with square symmetry, infinite in the horizontal direction and comprising $N$ 
periods in the vertical direction. The basic cell is given in fig. 2: it is made of square air holes (side $d/2$) inside a dielectric matrix of permittivity $\e=9$.
We first compute the transmitted field for $N=1$: the structure is made of one single grating. The incident
beam is a gaussian p-polarized beam ($w=50 d,\lambda /d=2.5,\theta =40^o$).
As it has been explained before the transmitted field is a sum of beams: the total field on the upper
interface is given in fig. 3 and the amplitudes of the different beams $\left|u^p_t(x,Nh) \right|$ for $p=0,1,2,3,4$ in fig. 4. It is clearly seen that there is a strong overlap between the beams. The shift of the first beam is 
$\Delta_n/d=1.66$ while the shift of the total transmitted beam is $\Delta_f/d=0.27$. The Bloch diagram is given in
fig. 5 (black curve). The value of $k_m$ is: $k_m\times d=2\pi/2.5 \times \sin(40^o)\sim 1.6$. The predicted shift is $\Delta_B/d=-1.36$, the renormalized Bloch diagram (grey curve in fig. 5) gives: $\Delta_R/d=1.66$. In this situation, the overlap between the multiple beams is strong and the transmitted field cannot be reduced to the first beam.

Let us now keep the same parameters except for the number of periods: we take $N=4$. This time the situation changes drastically: the transmitted field shows clearly a negative refraction (fig. 6), we have $\Delta_n/d \sim -36$, the shift of the total field is $\Delta_f/d \sim -32$ and the Bloch diagram gives: $\Delta_B/d\sim -11 $. In this situation the overlap between the beams is less important (fig. 7). The renormalized Bloch diagram gives: 
$\Delta_R/d\sim -27$. The discrepancy between $\Delta_n$ and $\Delta_R$ is due to the fact that the beam is not spectrally narrow enough. If we use a narrower beam by taking $w=200d$ we get a better result:  $\Delta_n/d \sim -28$. 
Let us look at the modulus
of the eigenvalues of ${\cal T}$ and the ratii on the different modes(fig. 8). On this figure, the modulus of the eigenvalues are given in thin solid line, the ratio over the propagative waves is shown in thick solid line and the ratio over the evanescent waves in dashed line. It is clearly seen that near $\lambda/d \sim 2.5$ there are more evanescent waves than propagative ones: this explains why the Bloch diagram gives a false description.
Let us pursue and take $N=16$, keeping the same parameters as for $N=4$, we get a transmitted field given in fig. 9 the beam is hugely deformed and the value of its barycenter is not relevant: here the behavior of the field is much more complicated than could be expected from the Bloch diagram.

Although the above results show that the prediction of the Bloch diagram can be quite false, it should be said that, provided the influence of the evanescent waves is not too strong, the Bloch diagram can give very accurate results. This is the case if we take $\lambda/d=6$ with the same parameters as above. We get:
$\Delta_n/d=6.55$ and $\Delta_B/d=6.45$. This time the Bloch diagram predicts fairly well the position of the first transmitted beam.

\section{Conclusion}
We have shown that the shift of the first transmitted beam
is given by the normal to the renormalized isofrequency diagram and that it can be quite different from the prediction of the Bloch diagram. We have proven that this effect is due to the existence of evanescent waves inside the structure. The analysis uses a dressed transfer matrix that could be extended to aperiodic or random structures \cite{graded,local}.
We have also pointed out the strong influence of the number of periods and of the spectral widths of the beams: these parameters can drastically modify the behavior of the beam. Moreover, it may happen that the total transmitted field be not properly described by the first transmitted beam. In that situation, neither the Bloch diagram nor the renormalized diagram can give reliable results.
This study aims at showing that one should be prudent when using the Bloch diagram to predict the behavior of the field inside a true (i.e. finite) photonic crystal. Beyond that, it shows that many new effects can be expected due to the presence of evanescent waves. These are not a drawback but rather represent new possibilities to imagine new fonctionnalities.

\textbf{Acknowledgments}\\
This work was realized in the framework of the ANR project POEM PNANO 06-0030.

\newpage

\newpage
Figures captions \\
Figure1: 2D Photonic crystal as stacks of gratings: periodic in x and finite in y.\\
Figure 2: Basic cell of the photonic crystal used in the numerical experiments.\\
Figure 3: Amplitude of the transmitted field on the upper interface.\\
Figure 4: Amplitude of the first five transmitted beams.\\
Figure 5: Black line: Bloch diagram for the photonic crystal. Gray line: renormalized Bloch diagram. $k_x\times d$ and $k_y \times d$ belong to $(-\pi,+\pi)$. 	The vertical dashed line indicates the average value of $k_x$ in the incident field. \\
Figure 6: Amplitude of the transmitted field on the upper interface.\\
Figure 7: Amplitude of the three first transmitted beams.\\
Figure 8: Thin solid line: modulus of the eigenvalues of the transfer matrix. Thick solid line (blue online): ratio of the field on the propagating modes. Dashed line (red online): ratio of the field on the evanescent modes. Dashed dotted line (green online): ratio of the field on the anti-evanescent modes. This arrows indicate the wavelengths were the computations were performed.\\
Figure 9: Amplitude of the transmitted field on the upper interface.\\

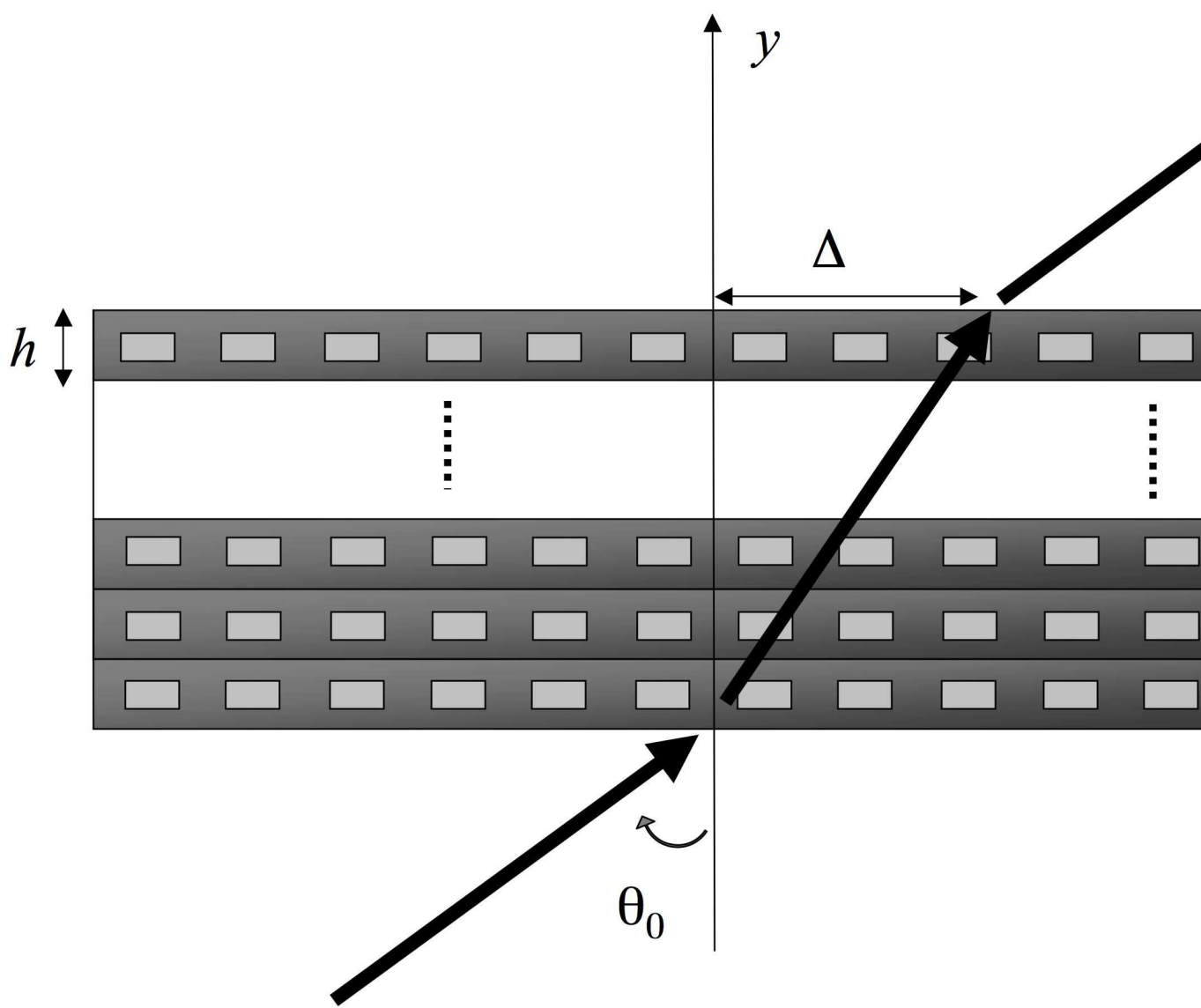
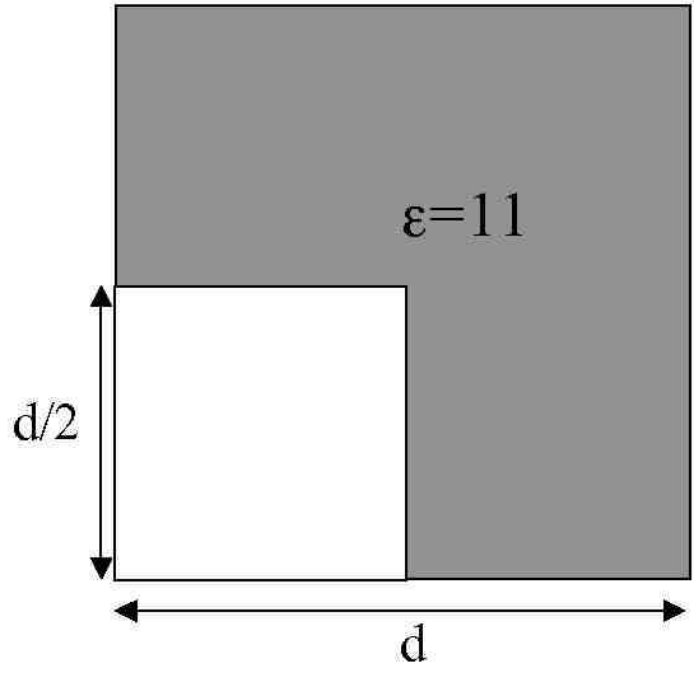
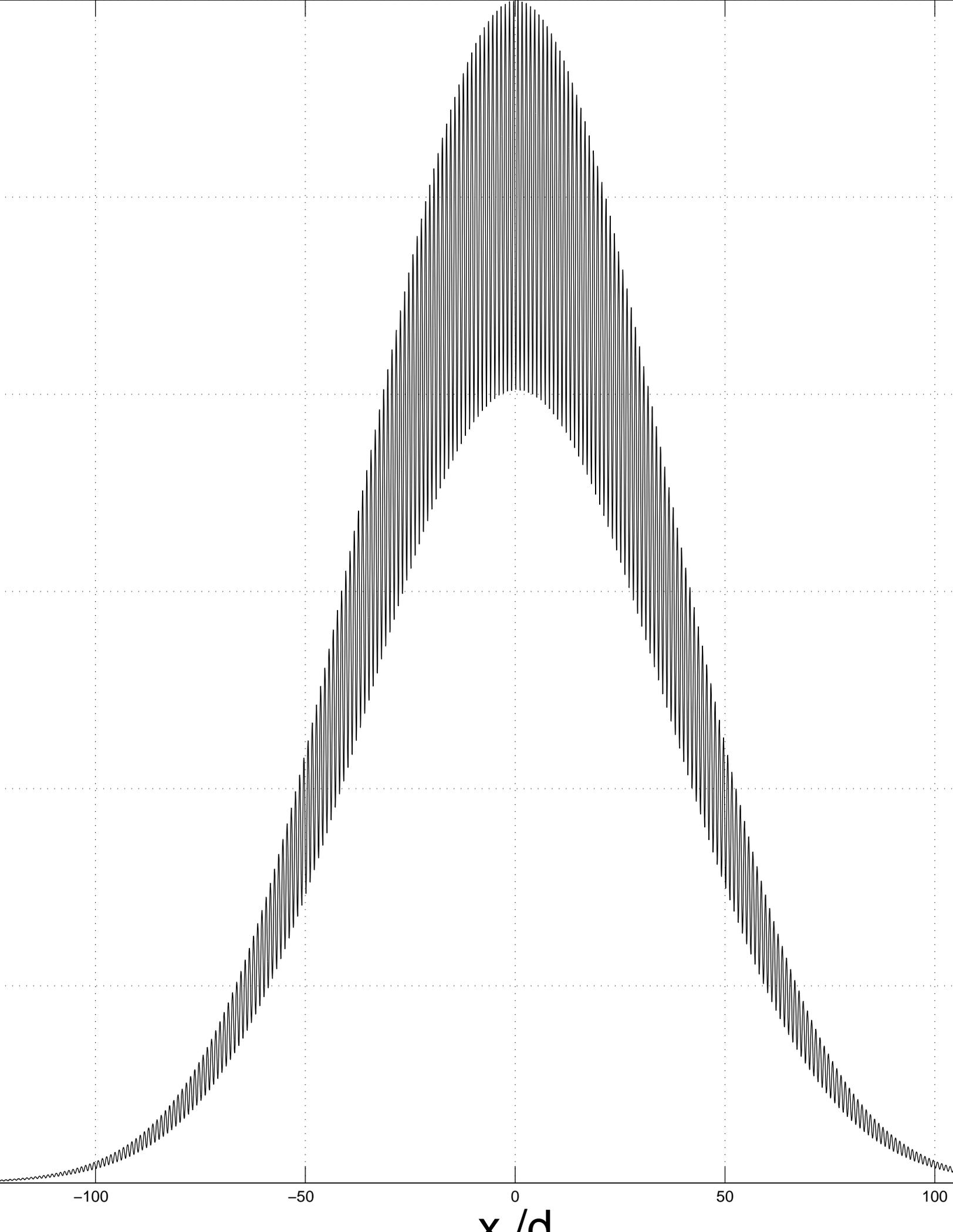
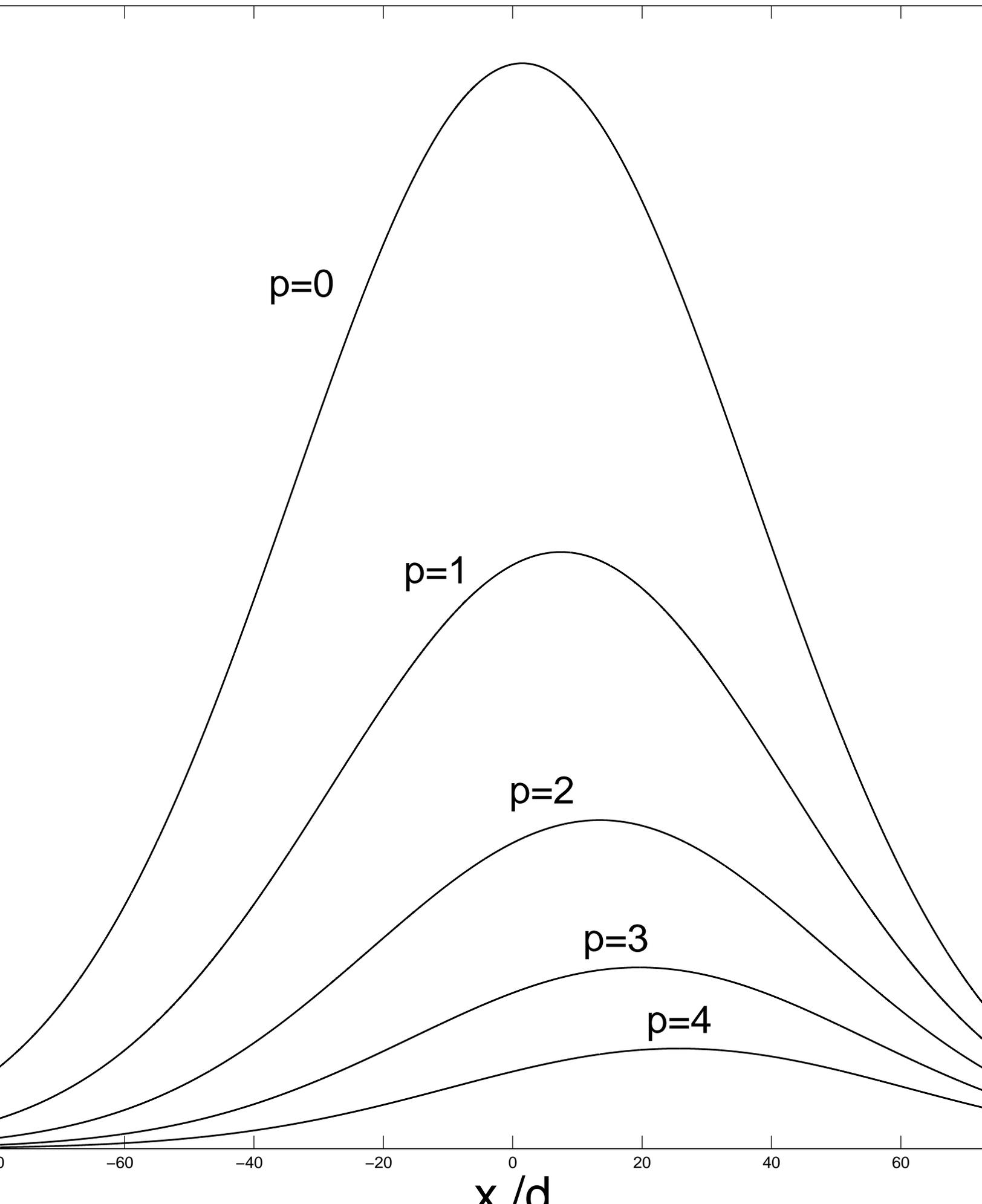
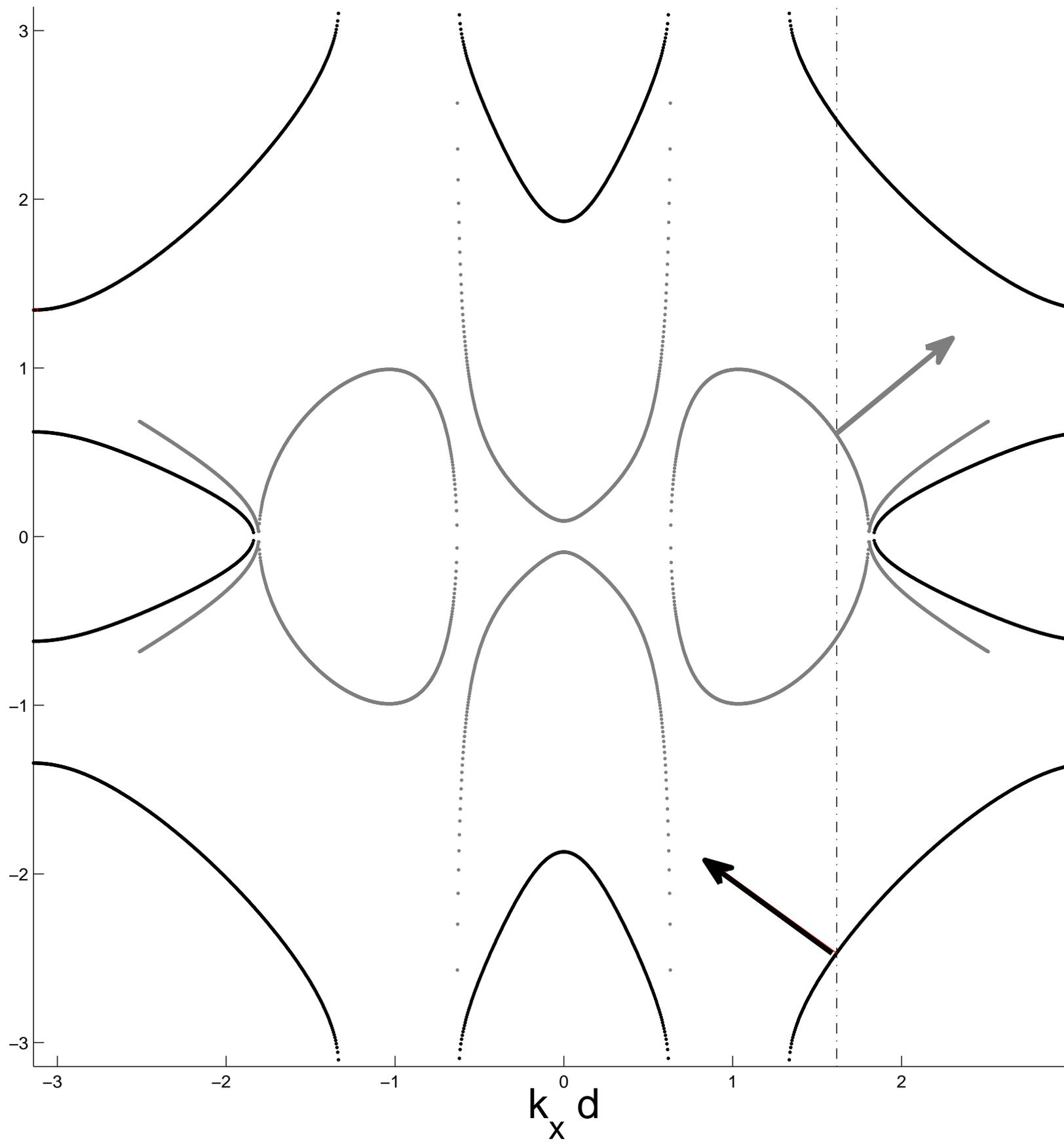
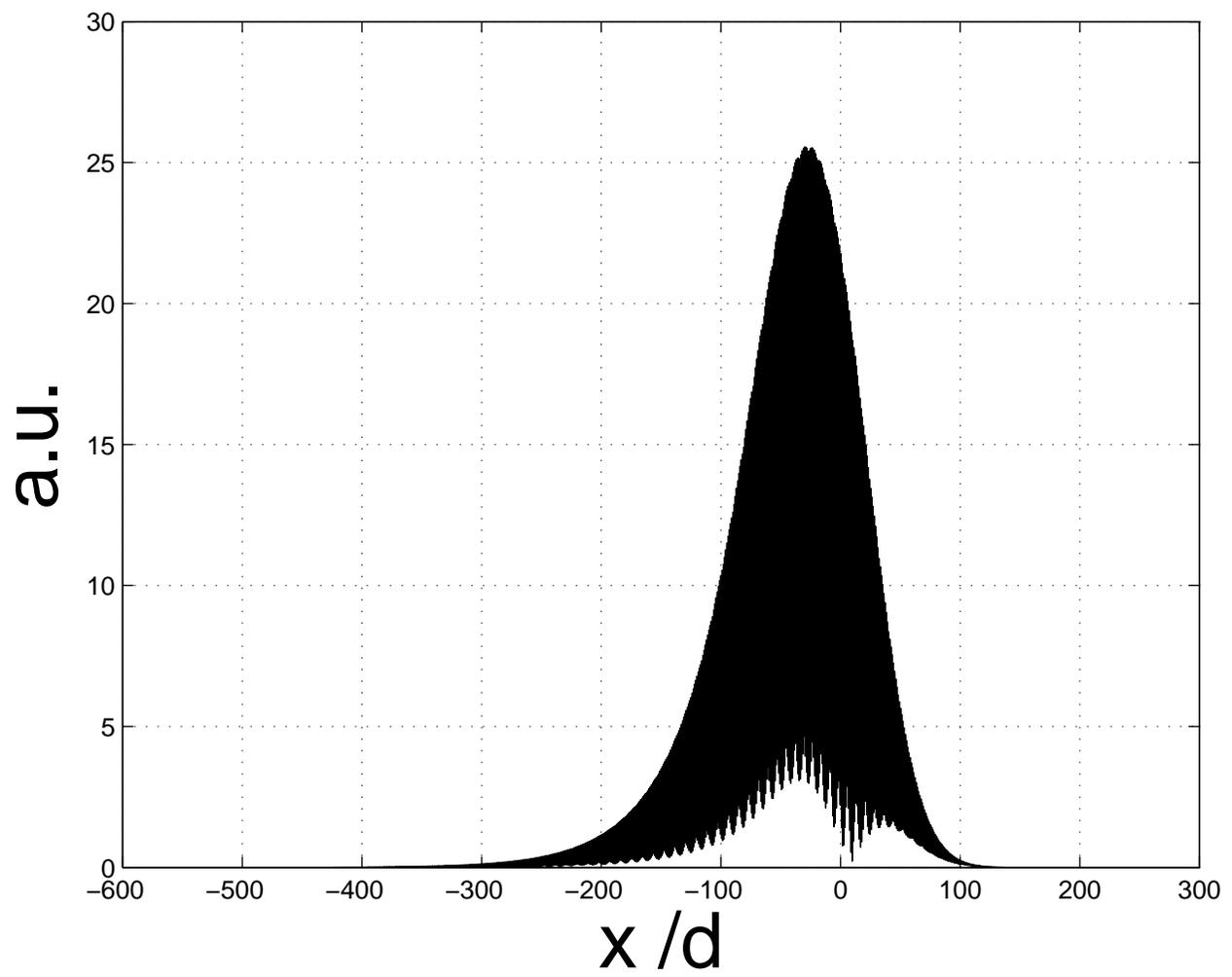
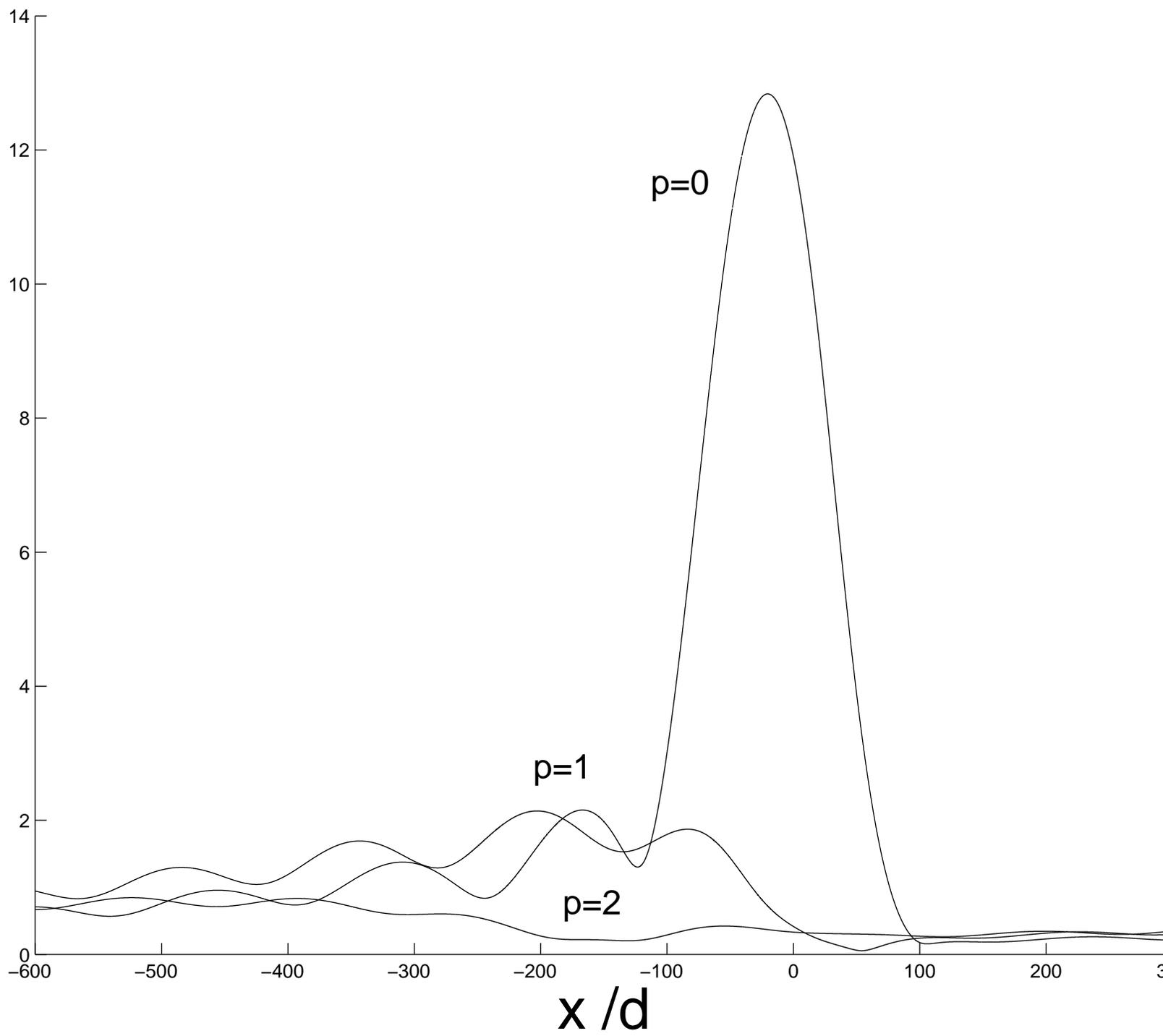
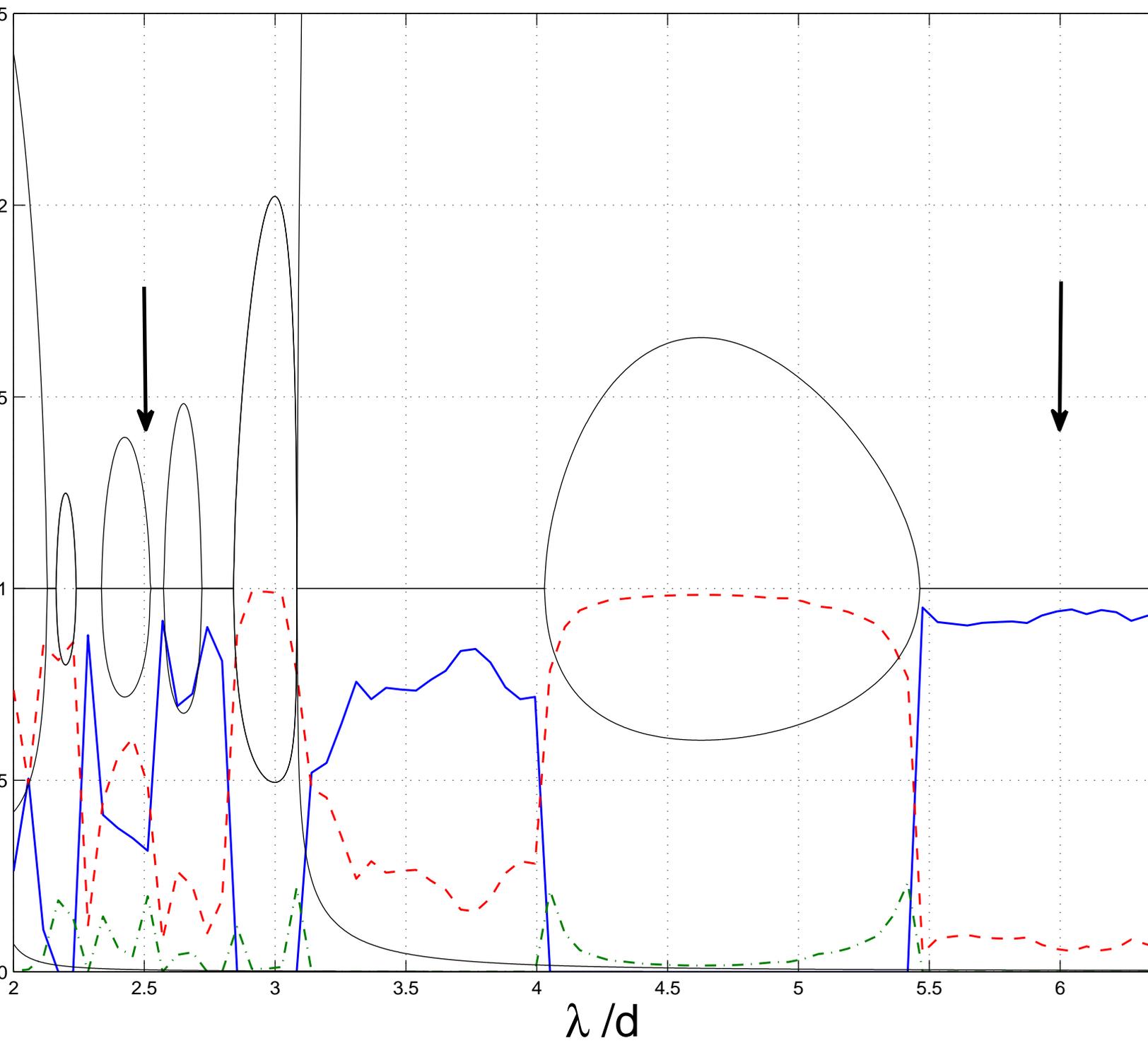
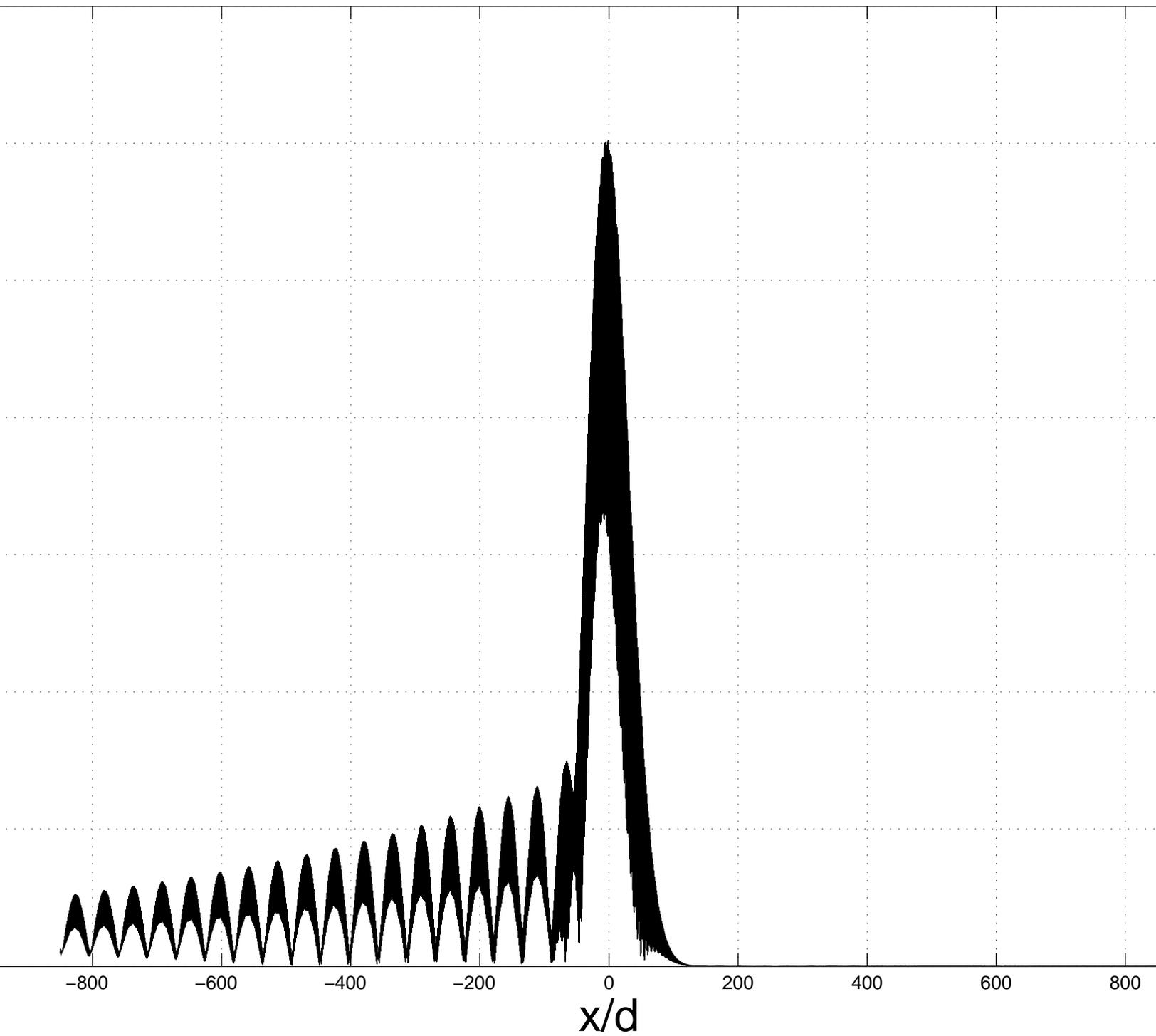
\end{document}